# Time relaxation of microwave second order response of superconductors in the critical state

A. Agliolo Gallitto, M. Guccione, and M. Li Vigni

*INFM and Dipartimento di Scienze Fisiche e Astronomiche, via Archirafi 36, I-90123, Palermo, Italy*

Relaxation of the microwave second order response of $YBa_2Cu_3O_7$ and $Ba_{0.6}K_{0.4}BiO_3$ crystals in the critical state is studied. The samples are exposed to static and pulsed microwave magnetic fields. The second harmonic signals decay during the time in which the microwave pulse endures. The decay times depend on the superconductor investigated and on the way the value of the static field has been reached.

## 1. INTRODUCTION

We report experimental results on microwave second harmonic (SH) emission in $YBa_2Cu_3O_7$ (YBCO) and $Ba_{0.6}K_{0.4}BiO_3$ (BKBO) crystals. The samples have been exposed to a static magnetic field, higher than the lower critical field, and an intense pulsed mw magnetic field. We have performed measurements of the SH signal intensity as a function of time, in the time interval in which the mw pulse endures. We have observed a signal decay that can be related to relaxation processes arising when fluxons are driven to move by em fields of high frequency.

## 2. EXPERIMENTAL RESULTS

The sample is placed in a bimodal cavity, resonating at two angular frequencies $\omega$ and $2\omega$, with $\omega/2\pi = 3$ GHz, in a region in which the mw magnetic fields $H(\omega)$ and $H(2\omega)$ are maximal and parallel to each other. The fundamental mode of the cavity is fed by a pulsed oscillator, with pulse repetition rate of 1 pps, pulse width of ~ 1 ms and maximal peak power of ~ 50 W. The SH signals radiated by the sample are detected by a superheterodyne receiver. All the measurements have been performed with the static magnetic field $H_0$ parallel to $H(\omega)$, at $T=4.2$ K.

Figure 1 shows the SH signal intensity as a function of the time, within the time interval of the mw pulse width, in YBCO (a) and BKBO (b) crystals. Measurements have been performed soon after $H_0$ had reached the value of 2 kOe, at increasing field (open circles) and at decreasing field (solid circles). All the curves are normalized to their maximal value.

The decay time is different for the two samples and depends on the way the $H_0$ value has been obtained. For both crystals the SH signal intensity decays faster when the value has been reached at increasing $H_0$. Further, the time dependence of the SH signal cannot be described by a single exponential law. Measurements performed at different $H_0$ values have shown that the decay time depends weakly on the intensity of the static field.

The figure shows that no decay is observed in YBCO at decreasing $H_0$. All the other curves are characterized by a first relatively rapid decay followed by a slower one. For YBCO at increasing field we have estimated the first decay time to be $\tau_1 \approx 1$ ms and the second one $\tau_2 \approx 4$ ms. For BKBO $\tau_1 \approx 0.1$ ms and $\tau_2 \approx 0.3$ ms at increasing $H_0$, and $\tau_1 \approx 0.3$ ms and $\tau_2 \approx 1$ ms at decreasing $H_0$. We remark that measurements in YBCO powder dispersed in polystyrene have shown a time dependence similar to that of fig.1 (a).

## 3. DISCUSSION

It is well known that superconductors (SC) in the critical state, when submitted to an intense em field, exhibit a magnetization vector containing Fourier components at harmonic frequencies of the driving field. The harmonic emission by type II SC in the critical state has been for the first time studied by Bean [1]. The Bean model describes quite well the response of the fluxon lattice to em fields of low frequency. It is tacitly assumed that the flux lattice follows the em field variations. In this case only odd harmonic emission is expected, in agreement with the results on conventional SC at frequencies of kHz [1]. However, it has been shown that SC in a critical state exposed to intense pulsed mw fields exhibit odd as well as even harmonic emission. In order to discuss these results, Ciccarello *et al.* [2] have elaborated a model in which a rectification process operated by SC in the critical state is hypothesized. Because of the rigidity of the fluxon lattice, the induction field inside the sample does not follow the variations of an em field of high frequency, except when the variation involves motion of fluxons in the surface layers of the sample. It has been supposed that, for "direct critical state" developed by increasing fields, the induction flux in the sample does not vary during the positive semi-period of the mw field, while it does during the negative semi-period. The opposite occurs when decreasing fields develop a "reverse critical state". On this hypothesis, the response of the sample will be uneven during the period of the microwave field, with consequent even harmonic emission.

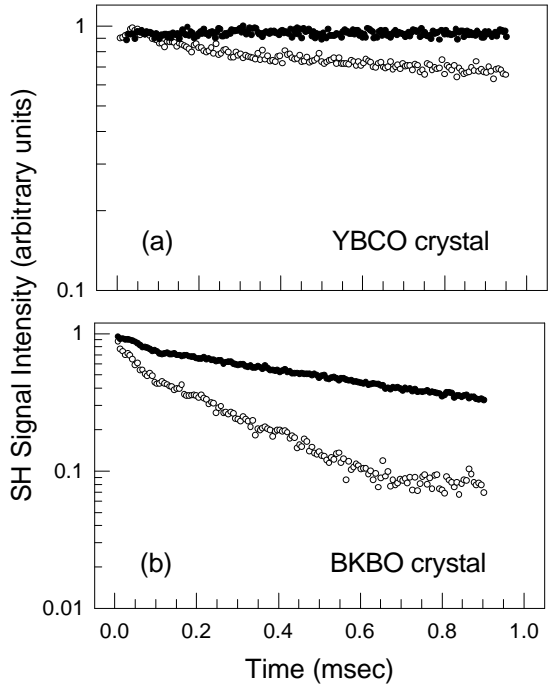

Figure 1. Time dependence of SH signal. $H_0 = 2$ kOe attained increasing (open circles) and decreasing (solid circles) the external field. $T$=4.2 K, input peak power ≈30 W.

In the model developed by Ciccarello *et al.* the drastic assumption is done that the fluxon lattice, being unable to follow the mw field variations, completely cuts out a semi-period of the mw field. In this case a stationary SH signal is expected. A different assumption can be done on supposing that the fluxon lattice tries to follow the mw field variations, by responding to a high frequency excitation with a characteristic time related to flux redistribution processes. Now the rectification process can be not complete, even harmonic emission is still expected and the SH signal is expected to decay during the time for which the microwave pulse endures.

The results reported in the figure seams to confirm that the latter hypothesis is verified, especially for the BKBO crystal. Further, our results show that the decay time is longer in the inverse than in the direct critical state.

A transient response, similar to that observed by us, has been reported in mw power absorption measurements in ceramic YBCO exposed to step-like changes of the static field [3]. The results refer to measurements performed at low static fields where processes occurring in Josephson fluxons come into play. The authors observed an initially rapid decay followed by a slower one. They ascribed the response to transient shielding currents reflecting processes of re-distribution of magnetic flux.

Our measurements have been performed at low temperatures and high magnetic fields; consequently, our results reflect relaxation processes of Abrikosov fluxons in the critical state. We would like to remark that at higher temperatures, where no critical state can be reasonably hypothesized, we have not detected any decay of the SH signals.

In conclusion, we have shown that SC in the critical state exhibit a transient second order mw response, which does not follow a single exponential law. The transient may be related to processes of re-distribution of magnetic flux, which came into play when fluxons are driven by em fields of high frequency. Our results show that the relaxation kinetics is slower when the fluxons are forced to leave the samples than when they are forced to penetrate into the samples.